\documentclass{cernrep} 
\usepackage[T1]{fontenc}
\usepackage{graphicx}

\providecommand{\prob}{\mathrm{P}}
\providecommand{\differential}{\mathrm{d}}

\begin{document}

  \title{Bayesian versus frequentist upper limits}
  \author{Christian R\"{o}ver$^1$, Chris Messenger$^{1,2}$ and Reinhard Prix$^1$}
  \institute{$^1$Max-Planck-Institut f\"{u}r Gravitationsphysik (Albert-Einstein-Institut), Hannover, Germany\\$^2$School of Physics \& Astronomy, Cardiff University, Cardiff, UK}
  \date{\today}

  \maketitle

\begin{abstract}
  While gravitational waves have not yet been measured directly, data
  analysis from detection experiments commonly includes an upper limit
  statement.  Such upper limits may be derived via a frequentist or
  Bayesian approach; the theo\-reti\-cal implications are very different,
  and on the technical side, one notable difference is that one case
  requires \textsl{maximization} of the likelihood function over
  parameter space, while the other requires
  \textsl{integration}. Using a simple example (detection of a
  sinusoidal signal in white Gaussian noise), we investigate the
  differences in performance and interpretation, and the effect of the
  ``trials factor'', or ``look-elsewhere effect''.
\end{abstract}

\section{Introduction}
\subsection{Upper limits}
In general, an upper limit is a probabilistic statement bounding one
of several unknown parameters determining the observed data at hand.
While it would be hard to derive general properties applicable in any
possible data analysis context, we will for the illustration purpose
consider a simple case here: a sinusoidal signal in white Gaussian
noise. This example exhibits many similarities with commonly
encountered real-world problems, including the use of Fourier methods,
nuisance parameters, trials factors, partly analytical and numerical
analysis, etc., and we believe is general enough to yield valuable
insights.

\subsection{The frequentist case}
The frequentist detection approach is based on some \textsl{detection
statistic}~$d$, which for given data is then used to derive a
significance statement along the lines of \textsl{``If the data were
only noise (null hypothesis~$H_0$), a detection statistic value~$\geq
d_0$ would have been observed with probability~$p$.''} ($\prob(d\geq
d_0\,|\,H_0)=p$).  The probability~$p$ here is the p-value, and a low
p-value is associated with a great significance. In the case of a
non-detection, the statement then may be reversed to an upper limit
statement \textsl{``Had the signal amplitude been $\geq A^\star$, a
larger detection statistic value ($\geq d_0$) would have been observed
with at least 90\% probability''} ($\prob(d\geq d_0\,|\, A\geq
A^\star)\geq 90\%$), where $A^\star$ is the 90\% confidence upper
limit (e.g.\ \cite{AbbottEtAl2004c,BradyCreightonWiseman2004}).

\subsection{The Bayesian case}
In the Bayesian framework, detection and parameter estimation are more
separate problems; for detection purposes one would need to derive the
\textsl{marginal likelihood}, or \textsl{Bayes factor}, which (in
conjunction with the prior probabilities for the ``signal'' and
``noise~only'' hypotheses $H_1$ and $H_0$) allows one to derive the
probability for the presence of a signal. The detection statement
would then be \textsl{``(Given the observed data~$y$,) the probability
for the presence of a signal is~$p$.''} ($\prob(H_1|y)=p$). The upper
limit statement on the other hand is a matter of parameter estimation;
given the joint posterior distribution of all unknowns in the model,
one would need to marginalize to get the posterior distribution of the
parameter of interest alone. The upper limit statement would then be
``\textsl{(Given the observed data and the presence of a signal,) the
amplitude is $\leq A^\star$ with 90\% probability.''}  
%
($\prob(A\leq A^\star\,|\, y,H_1)=90\%$) 
\cite{BDA,AbbottEtAl2010a}.

\section{The data model}
We assume the data~$y$ to be a time series given by a parameterized
signal~$s$ and additive noise~$n$:
\begin{equation}
  y(t_i) \; = \; s(t_i) + n(t_i),
\end{equation}
where $i=1,\ldots,N$ and $t_i=i\Delta_t$.
The (sinusoidal) signal is given by
\begin{equation}
  s(t) \;=\; A\,\sin(2\pi f t + \phi),
\end{equation}
where $A\geq0$ is the amplitude, $0\leq\phi<2\pi$ is the phase, and
$f \in \{\frac{j_1}{N\Delta_t},\ldots,\frac{j_k}{N\Delta_t}\}$ is
the frequency, where $1\leq j_1,\ldots,j_k \leq \frac{N}{2}-1$ defines
the range of possible (Fourier) frequencies.  
The number $k$ of frequency bins may be varied and constitutes the
so-called ``trials factor'' here.  The noise~$n$ is assumed to be
white and Gaussian with variance~$\sigma^2$.

\section{Frequentist approach}\label{sec:FreqentistApproach}
If there were no unknown parameters in the signal model, then,
following from the Neyman-Pearson lemma, the optimal detection
statistic would be given by the \textsl{likelihood ratio} of the two
hypotheses.  In the case that the hypotheses include unknowns
(composite hypotheses) as in our case, this is commonly treated using
the \textsl{generalized likelihood ratio} framework, that is, by
considering the ratio of \textsl{maximized} likelihoods, where
maximization is done over the unknown parameters \cite{MGB}.

In our case, we have a 3-dimensional parameter space under the signal
model.  The conditional likelihood for a given frequency may be
maximized analytically over phase and amplitude. The \emph{profile
likelihood} (maximized conditional likelihood for given frequency, as
a function of frequency) is eventually proportional to the time
series' periodogram. The generalized likelihood ratio detection
statistic then is given as the periodogram maximized over the
frequency range of interest:
\begin{equation}\label{eqn:DetStatDefinition}
 d^2 \;:=\; \max_j \, {\textstyle \frac{2}{N\sigma^2}} \big|\tilde{y}_j\big|^2
\end{equation}
where $\tilde{y}_j$ is the (complex valued) $j$th element of the
discretely Fourier transformed time series~$y$.  The
``$\frac{2}{N\sigma^2} \big|\tilde{y}_j\big|^2$'' term (the
periodogram) maximized over in~(\ref{eqn:DetStatDefinition}) is in
fact also the \textsl{matched filter} for a sinusoidal signal
\cite{WainsteinZubakov}, and the maximum~$d^2$ is commonly referred to
as the ``loudest event'' \cite{BradyCreightonWiseman2004}.

The detection statistic's distribution may be derived analytically
under both hypotheses~$H_0$ and $H_1$, as this is a particular case of
an \textsl{extreme value statistic} \cite{MGB}.  Under the null
hypothesis, $d^2$ is the maxi\-mum of $k$ independently
$\chi_2^2$-distributed random variables; the cumulative distribution
function (CDF) of~$d^2$ is given by
\begin{equation}\label{eqn:DetStatDistnH0}
  F_{d^2;H_0}(x) \;=\; \prob(d^2\leq x \,|\, H_0) \;=\; \bigl(F_{\chi_2^2}(x)\bigr)^k
\end{equation}
where $F_{\chi_2^2}$ is the CDF of a $\chi_2^2$ distribution, and $k$
again is the number of independent frequency bins, or ``trials''.
This is essentially the ``background distribution'' of~$d^2$.
Under the signal hypothesis~$H_1$, $d^2$~is the maximum of $(k\!-\!1)$
independently $\chi_2^2$-distributed random variables \emph{and} one
noncentral-$\chi_2^2(\lambda)$-distributed variable with noncentrality
parameter~$\lambda=\frac{N}{2\sigma^2}A^2$.  The corresponding CDF
under $H_1$ then is
\begin{equation}\label{eqn:DetStatDistnH1}
  F_{d^2;H_1}(x) \;=\; \bigl(F_{\chi_2^2}(x)\bigr)^{(k-1)} \times F_{\chi_{2,\lambda}^2}(x)
\end{equation}
where $F_{\chi_{2,\lambda}^2}$ is the CDF of a noncentral $\chi_2^2$
distribution with parameter~$\lambda$.

For some observed detection statistic value~$d^2_0$, the (detection)
significance is determined by the p\mbox{-}value $\prob(d^2 \geq d^2_0
\,|\,H_0) = \int_{d^2_0}^\infty p(d^2|H_0)\,\differential d^2$.  The
90\% loudest-event upper limit is given by the smallest amplitude
value~$A^\star$ for which $\int_{d^2_0}^\infty
p(d^2\,|\,A,H_1)\,\differential d^2 \geq 90\%$, so that $\prob(d^2\geq
d^2_0 \,|\, A \geq A^\star,H_1)\geq 90\%$.

\begin{figure}
  \centering
  \includegraphics[width=0.9\linewidth]{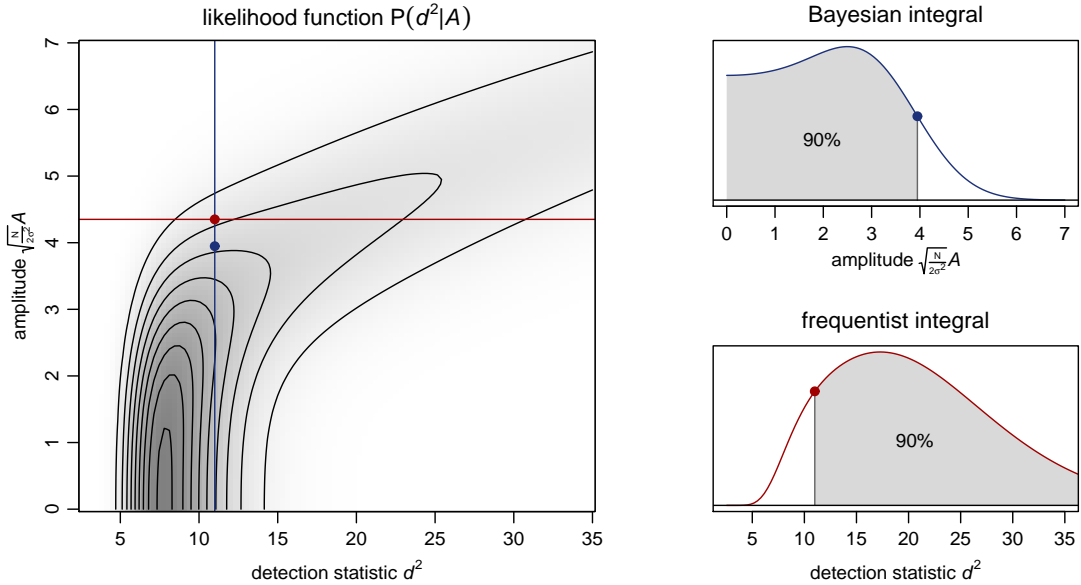}
  \caption{The integrals to be computed for a frequentist and a
           Bayesian 90\%~upper limit are very different. The Bayesian
           integral is computed along the vertical amplitude axis,
           conditioning on the observed detection statistic
           value~$d^2=d^2_0$.  The frequentist integral goes along the
           horizontal axis of possible realisations of~$d^2$ for any
           given amplitude.  (Example values here: $N=100$,
           $\Delta_t=1$, $\sigma^2=1$, $k=49$, $d^2_0=11$.)}
  \label{fig:integration}
\end{figure}

\section{Bayesian approach}
We assume uniform prior distributions on phase, frequency, and
amplitude. Given the (3-dimensional) likelihood function
\cite{Bretthorst}, one can then derive joint and marginal posterior
distributions $\prob(A,\phi,f \,|\, y)$ and $\prob(A|y)$.  However,
Monte Carlo simulations show that --- in this particular model --- the
amplitude's marginal posterior distribution is virtually unaffected by
whether one considers the complete data~$y$, or only the ``loudest
event''~$d^2$.  The essential information about the signal amplitude
is contained in that loudest event, and the marginal amplitude
posterior is dominated by the conditional distribution of the loudest
frequency bin.  We find that the main difference between the two kinds
of limits in this model is \textsl{not} due to maximization vs.\
integration of the posterior; in the following we will therefore
consider only the simpler, directly comparable, and more illustrative
case of a Bayesian loudest event limit based on $\prob(A|d^2)$ instead
of $\prob(A|y)$.

Our relevant observable now is the ``loudest event''~$d^2$.
The likelihood function $\prob(d^2|A)$ was defined
through~(\ref{eqn:DetStatDistnH1}) in the previous section.  The 90\%
upper limit on the amplitude is given by the amplitude~$A^\star$ for
which $\int_0^{A^\star}p(A\,|\,d^2,H_1) \, \differential A = 90\%$,
so that $\prob(A < A^\star\,|\,d^2,H_1)=90\%$.

\section{Comparison}
The likelihood function here is a function of two parameters: the
observable~$d^2$ and the amplitude parameter~$A$\@.  Since the
amplitude prior is assumed uniform, the posterior distribution is
simply proportional to the likelihood, which allows for a nice
comparison of both approaches.  \Fref{fig:integration} illustrates the
integrations performed for both the frequentist and the Bayesian upper
limits for some particular realisation~$d^2=d^2_0$.

\begin{figure}
  \centering
  \includegraphics[width=0.75\linewidth]{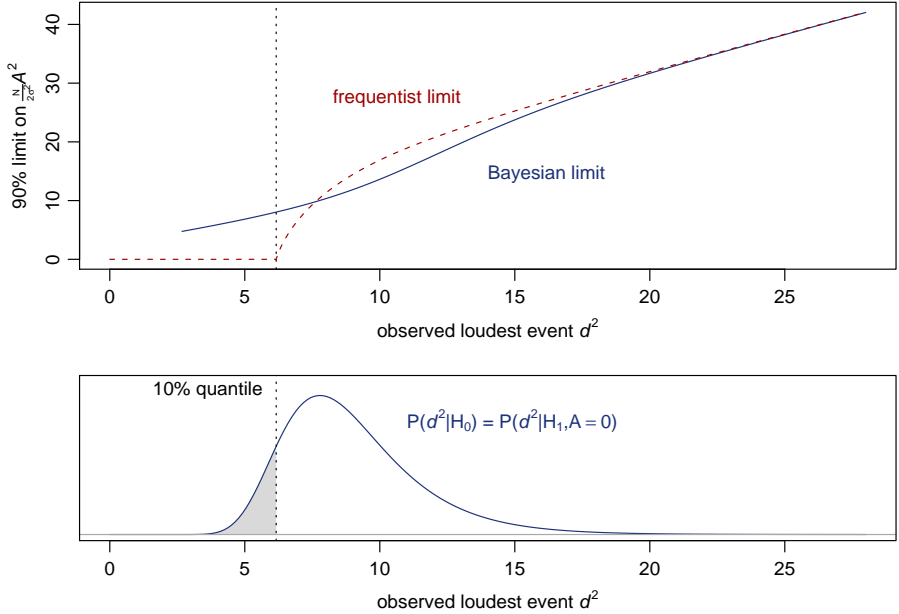}
  \caption{The mapping from observable~$d^2$ to the upper limit on amplitude.
           The bottom panel shows the ``background'' distribution of $d^2$ under $H_0$.
           (Example values here: $N=100$, $\Delta_t=1$, $\sigma^2=1$, $k=49$.)}
  \label{fig:mapping}
\end{figure}

Since the data~$y$ are reduced to a single observable~$d^2$, there
also is a one-to-one mapping from $d^2$ to the upper limit~$A^\star$.
\Fref{fig:mapping} shows both resulting upper limits as a function of
the ``loudest event''~$d^2$.
An important feature to note is that the frequentist limit will be
\textsl{zero} for certain values of~$d^2$. The point at (and below)
which this happens is the lower 10\% quantile of the 
%
``background'' distribution
of~$d^2$ under~$H_0$~(\ref{eqn:DetStatDistnH0}) --- at this point the
probability of observing a larger $d^2$ value is (by definition) 90\%
for zero-amplitude signals already, which makes zero the 90\% upper
limit.  Note that this implies that if $H_0$ in fact is true, 10\% of
all 90\% upper limits will be zero.  Note also that this is consistent
with the intended 90\% \textsl{coverage} of frequentist confidence
bounds --- if the upper limit is supposed to fall above and below the
true amplitude value with 90\% and 10\% probabilities respectively,
then 10\% of the upper limits \textsl{must} be zero under~$H_0$.

\begin{figure}
  \centering
  \includegraphics[width=0.95\linewidth]{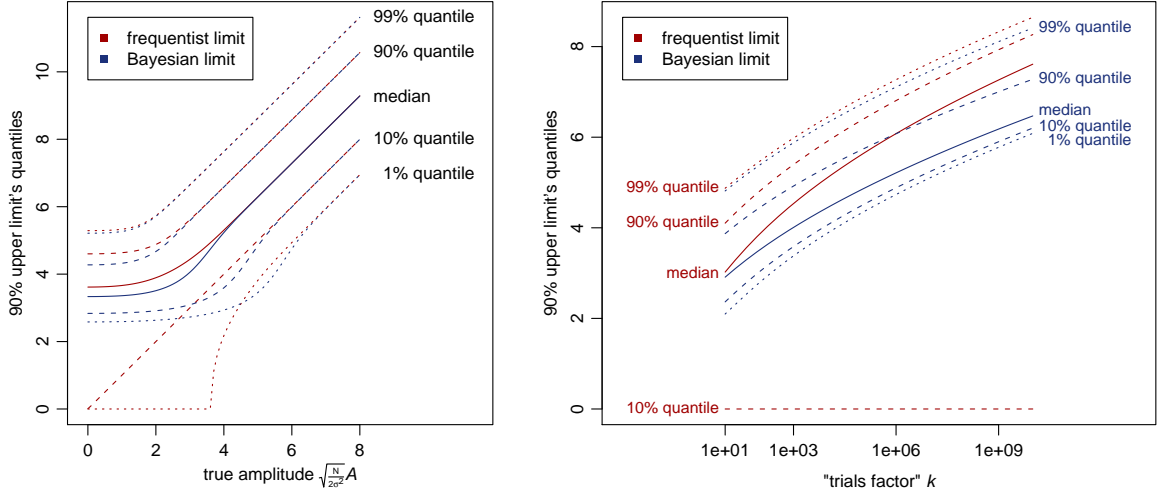}
  \caption{The distribution of upper limits as a function of amplitude
           (left panel) and trials factor (for zero amplitude; right
           panel).  Note that the frequentist 90\% limit is
           essentially a statistic that is designed to have its
           10\%~quantile at the true amplitude value.}
  \label{fig:limitdistns}
\end{figure}

Having the distribution of the detection statistic
(equations~(\ref{eqn:DetStatDistnH0}), (\ref{eqn:DetStatDistnH1})) and
the mapping from $d^2$ to upper limit (\Fref{fig:mapping}) allows us
to derive the distribution of upper limits for given parameters.
\Figure[b]~\ref{fig:limitdistns} illustrates the behaviour of the
resulting upper limits for different values of amplitude~$A$ and
trials factor~$k$.  The left panel shows that for large amplitudes the
two limits behave roughly the same, as one could already see from
\Fref{fig:mapping}, while for low amplitudes the posterior upper limit
will level off and will not rule out amplitude values below a certain
noise level. The frequentist limit's distribution on the other hand
reaches all the way down to zero, and in particular the 90\%~limit's
10\%~quantile follows a straight line of slope~1 and intercept~0 ---
the frequentist 90\%~limit is (by construction) essentially a
statistic that has its 10\%~quantile at the true amplitude value.

The right panel of \Fref{fig:limitdistns} shows the differing
behaviour of both limits as a function of the trials factor~$k$ when
the true amplitude is zero. The frequentist limit's 10\% quantile
remains at zero (the true value), while the posterior limit is bounded
away from zero but otherwise tends to yield tighter constraints on the
amplitude, especially for large~$k$.

\section{Conclusions}
The most obvious technical difference between frequentist vs.\
Bayesian upper limits is in maximization vs.\ integration over
parameter space. This, however, is not --- at least in the example
discussed here --- the primary origin of discrepancies between the
two. When founding \textsl{both} limits on maximization (i.e., the
``loudest event''), the behaviour of the Bayesian limit is affected
very little; so the crucial information about the signal amplitude is
in fact contained in the loudest event.  Both kinds of upper limits
behave very similarly for ``loud'' signals, i.e., a large
signal-to-noise ratio (SNR), but their differences become apparent in
the interesting case of (near\hbox{-}) zero amplitude signals.  While
the Bayesian upper limit expresses what amplitude values may be ruled
out with 90\% certainty based on the data (and model assumptions), the
frequentist upper confidence limit is defined solely through its
``coverage'' property. The frequentist 90\%~limit needs to end up
above and below the true amplitude value with 90\% and 10\%
probability respectively, which simply means that the frequentist
limit may be any random variable that has its 10\% quantile at the
true amplitude. This in particular implies that for a true amplitude
of $A=0$ the limit has a 10\% chance of being zero as well, and it
makes the frequentist limit very hard to actually interpret, not only
if it actually happens to turn out as zero.  When considering the
effect of the trials factor (or look-elsewhere effect) in the low-SNR
regime where both limits behave differently, the posterior-based limit
will usually yield tighter constraints especially for large trials
factors, but it will never be zero.

The Bayesian upper limit based on the amplitude's posterior
distribution will of course change with changing prior
assumptions. For simplicity, we assumed an (improper) uniform
amplitude prior here, but this should actually be a conservative
choice in some sense, for a realistic prior in the continuous
gravitational-wave context would in general be much more concentrated
towards low amplitude values (something like the --- also improper ---
prior with density $p(A)\propto \frac{1}{A^4}$).

Another question is how exactly one would do the actual computations
for a Bayesian upper limit in practice --- the frequentist upper
limits are usually not computed via direct analytical or numerical
integration of the likelihood, but the integral (see
\Fref{fig:integration}) is determined in a nonparametric fashion via
Monte Carlo integration and bootstrapping of the data. While the
frequentist limit requires finding the amplitude~$A^\star$ at which
the integral ($\prob(d^2>d^2_0\,|\,A=A^\star)$) yields the desired
confidence level, an analogous procedure to derive the Bayesian upper
limit would probably require Monte Carlo sampling of $\prob(d^2|A)$
across the range of all amplitudes~$A$ in order to then do the
integral in the orthogonal direction.

Further complications arise especially for the frequentist limit when
the signal model gets more complex.  The general procedure required
for the Bayesian upper limit is rather obvious --- determine the
marginal posterior distribution of amplitude~$\prob(A|y)$, then
determine the 90\%~quantile.  The frequentist procedure on the other
hand may run into major problems.  For example, if there are multiple
parameters affecting the signal's SNR, a ``loudest event'' might be
hard to define, or to translate into a constraint on the amplitude. As
there may not be a simple one-to-one connection between SNR and
amplitude parameter as in the present case, the ``loudest event'' may
not be the only relevant figure to constrain the signal amplitude.
The consideration of nuisance parameters is generally tricky in a frequentist
framework and may effectively suggest the use of a Bayesian procedure instead
\cite{Searle2008}.
Computation also becomes more complicated if the frequency parameter
is not restricted to (``independent'') Fourier frequencies.  Note that
the reasoning behind the generalized likelihood ratio approach (see
Sec.~\ref{sec:FreqentistApproach}) leading to the ``loudest event''
concept was very much an ad-hoc construction in the first place.

\begin{figure}
  \centering
  \includegraphics[width=0.75\linewidth]{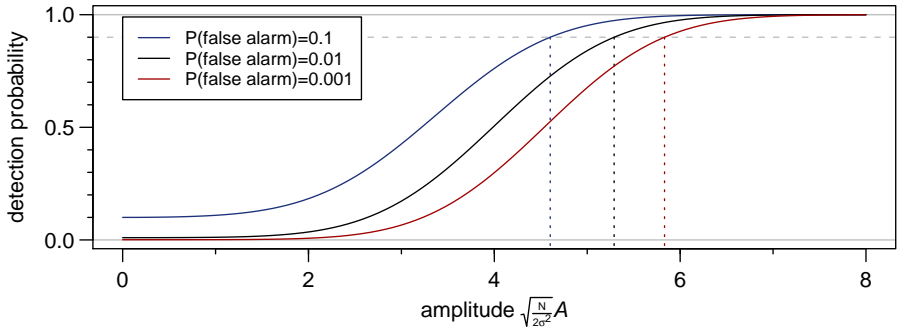}
  \caption{Illustration of the determination of a
           90\%~\textsl{detection sensitivity} threshold.  
           Such a statement would be independent of the observed data,
           and it requires the specification of an additional
           parameter: the corresponding false alarm rate defining the
           threshold of what is considered a ``detection''.  (Here:
           $N=100$, $\Delta_t=1$, $\sigma^2=1$, $k=49$.)}
  \label{fig:sensitivity}
\end{figure}

Another notable related concept is that of a \textsl{power constrained
upper limit}. In search experiments, these may be based on the
\textsl{sensitivity} of the search procedure. In case the search
yielded no detection, one can state the signal amplitude that would
have been detected with 90\%~probability; this number may then also be
used as a lower bound on the frequentist limit (\textsl{``don't rule
out what you wouldn't be able to detect''}). However, this kind of
statement requires the specification of another, additional parameter: the
corresponding false alarm rate defining the threshold of what is
considered a ``detection'', and as such is inseparably connected to
the detection procedure (see also \Fref{fig:sensitivity}). 
In particle physics a different approach is commonly taken; here the
sensitivity is usually specified as the expected upper limit for many
repetitions of the experiment in the absence of a signal. This figure
would correspond to the solid lines at zero amplitude in
Fig.~\ref{fig:limitdistns}.
An important point to note is that both these sensitivity statements 
do not depend on the observed data.
%
%

 \bibliographystyle{unsrt}
 \bibliography{../../literature/literature}

\end{document}